# Dirac Equation in the Presence of Hartmann and Ring-Shaped Oscillator Potentials

Z. Bakhshi*

Department of Physics, Faculty of Basic Sciences, Shahed University, Tehran, Iran.

July 9, 2018

## Abstract

The importance of the energy spectrum of bound states and their restrictions in quantum mechanics due to the different methods have been used for calculating and determining the limit of them. Comparison of Schrödinger-like equation obtained by Dirac equation with the non-relativistic solvable models is the most efficient methods. By this technique, the exact relativistic solutions of Dirac equation for Hartmann and Ring-Shaped oscillator potentials are accessible, when the scalar potential equals to the vector potential. Using solvable non-relativistic quantum mechanics systems as a basic model and considering the physical conditions provide the changes in the restrictions of relativistic parameters based on the non-relativistic definitions of parameters.

**Keywords:** Dirac equation; non-relativistic models; relativistic solutions; Hartmann potentials; Ring-Shaped oscillator potentials; scalar and vector potentials.

**PACS numbers: 03.65.w, 03.65.Fd 03.65.Ge,11.30.Pb**

*Corresponding author (E-mail: z.bakhshi@shahed.ac.ir)

# 1 Introduction

Since the advent of quantum mechanics, several methods have been developed in order to find the exact energy spectrum of bound states in stationary quantum systems. The knowledge of these spectrum is necessary for several applications in many fields of physics and theoretical chemistry [1-4]. Such encouraging results have arisen some studies on the potential within the frame work of common wave equations of both non-relativistic and relativistic wave equations i. e. including Schrödinger, Duffin-Kemmer-Petiau (DKP), Klein-Gordon or Dirac equations [5-10]. There are some non-central separable potential in spherical coordinates that are of considerable interest and practical in the branches of sciences such as chemistry and nuclear physics. Hartmann potential introduced by Hartmann is one of the non-central potential, which can be realized by adding a potential proportional to Coulomb potential [11-16]. This potential was suggested to describe the energy spectrum of Ring-Shaped potential obtained by replacing the Coulomb part of Hartmann potential with a Harmonic Oscillator term and that is called as a Ring-Shaped oscillator potential which is investigated to find discrete spectrum and integrals of motions [17-22]. The relativistic linear interaction, which is called the relativistic oscillator due to the similarity with the non-relativistic harmonic oscillator, has been subject of many successful theoretical studies. Such a space has interesting property and algebra for example some article in which a free particle has been studied in different situations, Dirac oscillator is system which is constituted by a relativistic fermion that is subjected to linear vector potential [23-25]. In this article, for solving Dirac equation with Hartmann and Ring-Shaped Oscillator potentials in three dimensions, assuming equality of scalar and vector potentials can be constituted the couple of differential equations for the spinor components [26-27]. One of them is the second-order differential equation for the upper spinor and the lower spinor can be gotten from the first-order differential equation based on the upper spinor. Since Hartmann and Ring-Shaped Oscillator potentials are contained two parts of radial and angular, the second-order differential equation is considered in the spherical polar coordinates. Separating of the second-order differential equation, there are two Schrödinger-like equations in $r$ and $\theta$ coordinates. Moreover, the normalized solution of the polar angular part is considered as an exponential function based on $\varphi$ coordinate and separating constant, because of there is not any part of $\varphi$ coordinate in the potential function. There exist one-dimensional solvable Schrödinger equation in the non-relativistic quantum mechanics for the determined potential which can be expanded to the Schrödinger-like equation is derived from Dirac equation [28-29]. In the radial part of differential equation, the relativistic energy spectrum can be gotten by comparing with the non-relativistic solvable Schrödinger equation. In this comparison, the relativistic energy spectrum is obtained based on the non-relativistic energy spectrum and the wave function of the non-relativistic space will be considered for calculation of the relations between non-relativistic and relativistic parameters. The mentioned method can be used on the angular part of differential equation. The relations of parameters between the two models is confirmed to the changes in the restriction of parameters. The new restriction of parameters and separating constants ensure the physical conditions. The paper is organized as: assuming $V(\vec{r}) = S(\vec{r})$, the couple of differential equation can be obtained for the spinor components and the second-order differential equation can be separated to the three coordinates in the 2 section. The radial part of Dirac equation and the relativistic energy spectrum that is associated with the radial part have been investigated in the 3 section. The angular part of Dirac equation for the potential that is related to $\theta$ coordinate according to different function of $\theta$, and the relativistic parameters have been paid attention in the 4 section. Finally, in the 5 section, the brief of method has been presented.

## 2    The general form of Hartmann and Ring-Shaped Oscillator potentials in Dirac equation

The generalized Hartmann and Ring-Shaped oscillator potentials are defined as follows [27]:

$$V(r,\theta) = V(r) + \frac{f(\theta)}{2r^2}. \tag{2.1}$$

The radial part of potential can be considered as coulomb and harmonic oscillator potentials [17-18]:

$$V(r) = -\frac{1}{2}(\frac{V_0\lambda}{r}), \qquad V(r) = Kr^2, \tag{2.2}$$

where $V_0$ , $\lambda$ and $K$ are free parameters with respect to the relevant potentials. Different types of functions are assumed for the angular part of potential so that the exact solvable models can be provided from Dirac equation. In Eq.(2.1), $\theta$ and $r$ are polar angular and radial in spherical coordinates of Hydrogen atom.

Time-independent Dirac equation for arbitrary scalar and vector potentials is given by differential equation:

$$\left[c\vec{\alpha}.\vec{P} + \beta(Mc^2 + \vec{S}(\vec{r}))\right]\psi(\vec{r}) = [\varepsilon - V(\vec{r})]\psi(\vec{r}). \tag{2.3}$$

The following parameters definitions satisfy in Eq.(2.3):

$$\vec{P} = -i\hbar\vec{\nabla}, \qquad \alpha \equiv \begin{pmatrix} 0 & \vec{\sigma} \\ \vec{\sigma} & 0 \end{pmatrix}, \qquad \beta \equiv \begin{pmatrix} I & 0 \\ 0 & -I \end{pmatrix}, \tag{2.4}$$

$\vec{\sigma}$ and $I$ are vector Pauli spin matrix and the identity matrix, respectively. Using the Pauli-Dirac representation as:

$$\psi(\vec{r}) = \begin{pmatrix} \varphi(\vec{r}) \\ \chi(\vec{r}) \end{pmatrix}, \tag{2.5}$$

where are spinor components. The following set of coupled equations for the spinor components can be gotten:

$$c\vec{\sigma}.\vec{P}\chi(\vec{r}) = \left[\varepsilon - V(\vec{r}) - Mc^2 - S(\vec{r})\right]\varphi(\vec{r}), \qquad (2.6)$$

$$c\vec{\sigma}.\vec{P}\varphi(\vec{r}) = \left[\varepsilon - V(\vec{r}) + Mc^2 + S(\vec{r})\right]\chi(\vec{r}). \qquad (2.7)$$

Assuming $S(\vec{r}) = V(\vec{r})$ and $S(\vec{r}) = -V(\vec{r})$ due to combine two Eqs.(2.6) and (2.7) and provide the situations for obtain the second-order differential equations according to one of the component so that another component can be gotten by using the first-differential equation based on the determined component. Since in the case $S(\vec{r}) = -V(\vec{r})$ the treatment two Eqs.(2.6) and (2.7) is quite equivalent to the case $S(\vec{r}) = V(\vec{r})$, therefore the case $S(\vec{r}) = V(\vec{r})$ are considered and then the results of that case will be expanded to the second case [26-27].

The state $S(\vec{r}) = V(\vec{r})$ allows to make two differential equations for each component:

$$\chi(\vec{r}) = \left[\frac{c\vec{\sigma}.\vec{P}}{\varepsilon + Mc^2}\right]\varphi(\vec{r}), \qquad (2.8)$$

$$\left[c^2\vec{P}^2 + 2(\varepsilon + Mc^2)V(\vec{r})\right]\varphi(\vec{r}) = \left[\varepsilon^2 - M^2c^4\right]\varphi(\vec{r}). \qquad (2.9)$$

Schrödinger-like equation is obtained for the component $\varphi(\vec{r})$ by considering the definitions of $\vec{P}$ and $V(\vec{r})$ in Eq.(2.9):

$$\left[-\hbar^2c^2\vec{\nabla}^2 - (\varepsilon + Mc^2)\left(\frac{V_0\lambda}{r} - \frac{f(\theta)}{r^2}\right)\right]\varphi(\vec{r}) = \left[\varepsilon^2 - M^2c^4\right]\varphi(\vec{r}). \qquad (2.10)$$

Assuming a solution as:

$$\varphi(\vec{r}) = \frac{1}{r}u(r)\Theta(\theta)\Phi(\varphi), \qquad (2.11)$$

Eq.(2.10) can be separated to three differential equations in the three dimensions $\varphi$, $r$ and $\theta$ :

$$\frac{1}{\Phi}\frac{d^2\Phi}{d\varphi^2} = -m^2, \qquad (2.12)$$

$$-\frac{d^2u(r)}{dr^2} + \left[\frac{\rho}{r^2} + 2\frac{(\varepsilon + Mc^2)}{\hbar^2c^2}V(r)\right]u(r) = \frac{(\varepsilon^2 - M^2c^4)}{\hbar^2c^2}u(r), \qquad (2.13)$$

$$\frac{1}{\sin\theta}\frac{d}{d\theta}\left(\sin\theta\frac{d\Theta(\theta)}{d\theta}\right) - \left[\frac{m^2}{\sin^2\theta} + (\varepsilon + M^2c^4)f(\theta) - s\right]\Theta(\theta) = 0, \qquad (2.14)$$

where $m$ and $\frac{\rho}{r^2}$ are separation factors.

The normalized solution of Eq.(2.12) that satisfies the boundary conditions becomes:

$$\Phi(\varphi) = -\frac{1}{\sqrt{2\pi}}e^{im\varphi}, \qquad m = 0, \pm 1, \pm 2, \dots. \qquad (2.15)$$

# 3    The radial part solutions of Dirac equation

In this section, radial part of wave function Eq.(2.13) will be analyzed by corresponding to Generalized Laguerre differential equation. Coulomb and Harmonic Oscillator potentials are two potentials that are considered in Eq.(2.13), respectively. For these potentials, Eq.(2.13) can be converted to Generalized Laguerre differential equation with exact solution of Generalized Laguerre polynomials. In first case, substituting the radial part of potential as Coulomb potential in Eq.(2.13) [26-27]:

$$\frac{d^2u(r)}{dr^2} + \left[\frac{(\varepsilon^2 - M^2c^4)}{\hbar^2c^2} - \frac{\rho}{r^2} + (\frac{\varepsilon + Mc^2}{\hbar^2c^2})\frac{V_0\lambda}{r}\right]u(r) = 0, \qquad (3.1)$$

and considering units system ($\hbar = 2m = 1$), Eq.(3.1) can be compared to the following non-relativistic solvable model [28-29]:

$$\frac{d^2u_{n,l}(r)}{dr^2} + (E - V(r))u_{n,l}(r) = 0. \qquad (3.2)$$

Indeed, comparing radial Schrödinger-like equation to non-relativistic Schrödinger equation according to Coulomb potential with exact solution based on Generalized Laguerre polynomials, the results of non-relativistic can be expanded to relativistic models. Non-relativistic model for Coulomb potential has the following form:

$$\frac{d^2u(r)}{dr^2} + \left[ -\frac{e^4}{4(n+l+1)} - \frac{l(l+1)}{r^2} + \frac{e^2}{r} \right] u(r) = 0. \tag{3.3}$$

Therefore, relativistic parameters can be connected to non-relativistic parameters as follows:

$$\rho = l(l+1), \tag{3.4}$$

$$\left( \frac{\varepsilon + Mc^2}{c^2} \right) V_0 \lambda = e^2, \tag{3.5}$$

$$\frac{\varepsilon^2 - M^2c^4}{c^2} = -\frac{e^4}{4(n+l+1)^2}. \tag{3.6}$$

Since $\frac{e^2c^2}{4(n+l+1)^2} > 0$, relation of parameters (3.5) causes the condition $\mid \varepsilon \mid < M^2c^2$. Relativistic energy can be calculated based on defined parameters in non-relativistic solvable model by Combining above relations of parameters. Assuming $\tau = \frac{V_0\lambda}{2c(n+l+1)}$, relativistic energy can be obtained as follows [26-27]:

$$\varepsilon = Mc^2 \frac{1 - \tau^2}{1 + \tau^2}. \tag{3.7}$$

In non-relativistic model, the exact solution is considered for Eq.(3.2) as [28-29]:

$$u_{n,l}(r) = f(r)F(g(r)), \tag{3.8}$$

where F(g(r)) is a special function based on the internal function $g(r)$. Generalized Laguerre polynomials is an orthogonal polynomials that satisfies in Eq.(3.3). Therefore, that function can be expanded to Schrödinger-like equation (3.1) of radial part. Since $\alpha > -1$ in Generalized Laguerre polynomials $L_n^\alpha(g(r))$ and $\alpha = 2l + 1$ in non-relativistic model, condition of $l > -1$ satisfies in Generalized Laguerre polynomials. Therefore, according to relation of $\alpha = 2l + 1$ in relativistic model, $\rho < 0$ and $\rho > 0$ are considered for $-1 < l < 0$ and $l > 0$, respectively. In the last angular part section, it will be shown that relativistic energy are calculated based on non-relativistic energy and term of $\rho + \frac{1}{4}$. Since the sign of non-relativistic energy term is cleared, therefore, determining term of $\rho + \frac{1}{4}$ is very important because of the condition $\mid \varepsilon \mid < M^2c^2$. The term of $\rho + \frac{1}{4}$ should be signed for defined different $l$ parameter. According to parametric relation of $\rho = l(l+1)$, $\rho + \frac{1}{4}$ will be positive for each $l$ that is defined in the problem. It means that the sign of term $\rho = l(l+1)$ separated the limit of $l$ parameter. The $\rho$ relativistic parameter will be restricted by $\rho \geq 0$ for $l > 0$, and $-\frac{1}{4} \leq \rho \leq 0$ for $-1 < l < 0$. Since there is term of $n + l + 1$ in energy spectrum and for $n = l + 1$ singularity happens in the wave function, so in Generalized Laguerre polynomials that is related to differential equation (3.1), parameter $n$ is transformed to $n - l - 1$. Thus energy spectrum will be restricted and the problem of singularity will be disappeared. If the following non-normalized wave function is associated to differential equation (3.3) for internal function $g(r) = \left(\frac{e^2}{n+l+1}r\right)$ [28-29]:

$$u_{n,l}(r) \propto g^{(l+1)} exp(-\frac{g}{2}) L_n^{2l+1}(g(r)), \qquad (3.9)$$

the radial wave function that is considered to differential equation (3.1) as follows:

$$u_{n,l}^{(1)}(r) \propto (2kr)^{l+1} exp(-kr) L_{n-l-1}^{2l+1}(2kr), \qquad (3.10)$$

where $k = \frac{e^2}{2(n+l+1)}$ and $0 < r < +\infty$. The wave function that is satisfied in Schrödinger-like equation must be physically acceptable. Physical wave functions which is satisfied in the usual square-integrability condition as $\int_{x_1}^{x_2} \mid \Psi_n(x) \mid^2 dx < \infty$ for energy bound state, to ensure Hermiticity of Hamiltonian in Hilbert space spanned by its eigenfunctions. Since this integral must be finite, the wave functions have to a constant value or zero at the endpoints of definition internal of $\vec{V}$ potential. Therefore, solutions of Schrödinger-like equation should be checked at the endpoints of $[x_1, x_2]$ interval providing of square-integrability condition and investigation of physical situations in the wave functions. It is seen that wave function (3.10) is a square-integrability function at the endpoints of $[0, +\infty]$ interval, so that $u_{n,l}(r) \to 0$ when $r \to 0$ and $r \to +\infty$ for the range $l > 0$ and $-1 < l < 0$. Thus it will be physically acceptable wave function in restriction of $l$ parameter.

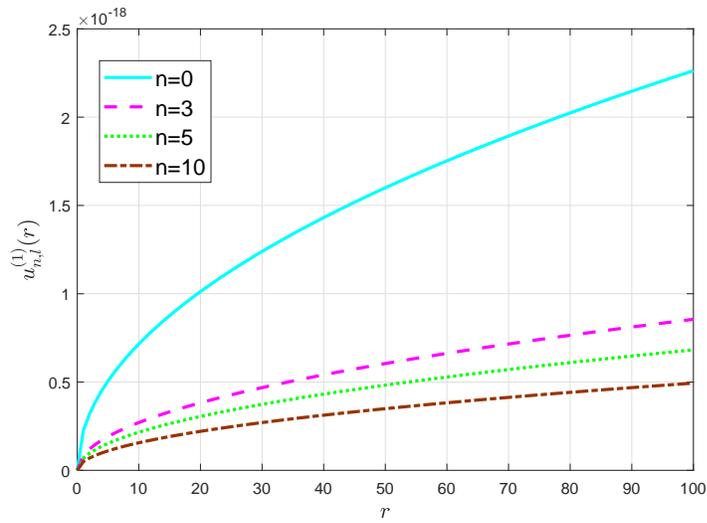

Figure 1: $u_{n,l}^{(1)}(r)$ versus $r$ with $l = -0.5$.

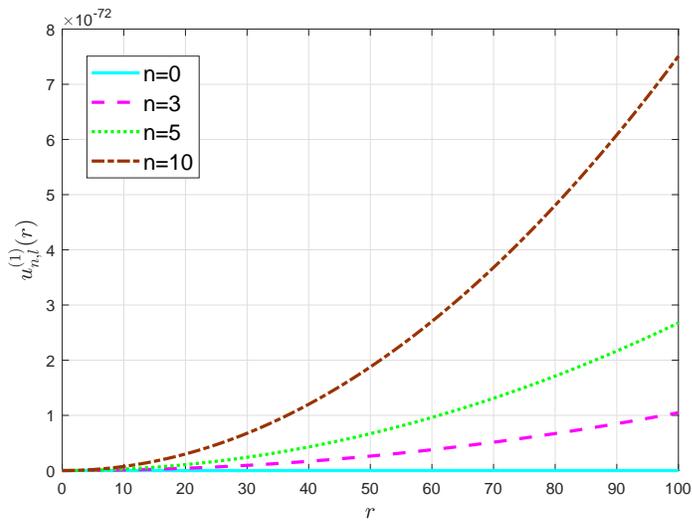

Figure 2: $u_{n,l}^{(1)}(r)$ versus $r$ with $l = 1$.

In the second case, the following differential equation is obtained from Eq.(2.13) for Harmonic Oscillator as radial part of potential:

$$\frac{d^2u(r)}{dr^2} + \left[ \frac{(\varepsilon^2 - M^2c^4)}{\hbar^2c^2} - \frac{\rho}{r^2} - (\frac{\varepsilon + Mc^2}{\hbar^2c^2})(2kr^2) \right] u(r) = 0, \qquad (3.11)$$

where $k > 0$. Non-relativistic solvable model based on Eq.(3.2) that can be compared to Eq.(3.11) has the following form ($\hbar = 2m = 1$) [28-29]:

$$\frac{d^2u(r)}{dr^2} + \left[2n\omega + (l+\frac{3}{2})\omega - \frac{l(l+1)}{r^2} - \frac{1}{4}\omega^2 r^2\right]u(r) = 0, \qquad (3.12)$$

where $\omega > 0$. The relations of parameters between Eq.(3.11) and Eq.(3.12) as follows:

$$\rho = l(l+1), \qquad (3.13)$$

$$\frac{2k(\varepsilon + Mc^2)}{c^2} = \frac{1}{4}\omega^2, \qquad (3.14)$$

$$\frac{\varepsilon^2 - M^2c^4}{c^2} = 2n\omega + (l+\frac{3}{2})\omega. \qquad (3.15)$$

Since $L_n^\alpha(g(r))$ Generalized Laguerre polynomials for $\alpha > -1$ is related to Eq.(3.12) as an exact solution and $\alpha$ parameter is defined as $\alpha = l + \frac{1}{2}$, therefore, $l$ parameter will be restricted by $l > -\frac{3}{2}$. The relations of parameters (3.13) and (3.15) emphasize to conditions of $\rho + \frac{1}{4} \geq 0$ and $\mid \varepsilon \mid < Mc^2$. Relativistic energy that is related to non-relativistic energy for Harmonic Oscillator can be gotten by combining relations (3.14) and (3.15) as:

$$(\varepsilon - Mc^2)^2(\varepsilon + Mc^2) = 8kc^2(2n + l + \frac{3}{2}), \qquad (3.16)$$

where is a third-order equation of $\varepsilon$. In non-relativistic solvable model, the wave function that is associated with Eq.(3.12) is [28-29]:

$$u_{n,l}(r) \propto g^{\frac{(l+1)}{2}} exp(-\frac{g}{2})L_n^{(l+\frac{1}{2})}(g(r)), \qquad (3.17)$$

where $g(r) = \frac{1}{2}\omega r^2$. By comparing two non-relativistic and relativistic models, the wave function (3.17) can be expanded to Eq.(3.11). Therefore, radial part of spinor wave function that can be corresponded to Eq.(3.11) as follows:

$$u_{n,l}^{(2)}(r) \propto \left(\frac{1}{2}\omega r^2\right)^{\frac{(l+1)}{2}} exp(-\frac{1}{4}\omega r^2)L_n^{(l+\frac{1}{2})}(\frac{1}{2}\omega r^2), \qquad (3.18)$$

where $-\infty < r < +\infty$. In the investigation of square-integrability condition, it is obvious that wave function (3.18) is limited as $u_{n,l}(r) \to 0$ when $r \to -\infty$ and $r \to +\infty$ in restriction of $l$ parameter that has been introduced as $l > -\frac{3}{2}$.

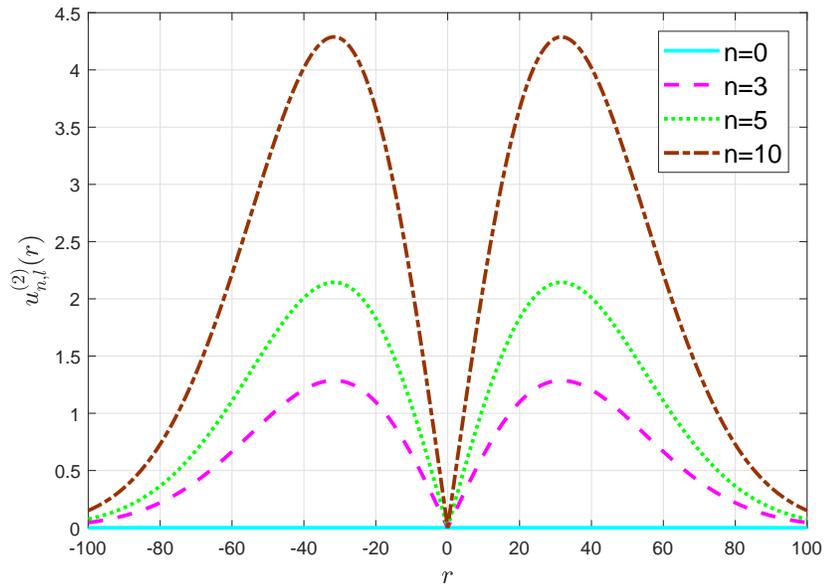

Figure 3: $u_{n,l}^{(2)}(r)$ versus $r$ with $l = 0$ and $\omega = 10^{-3}$.

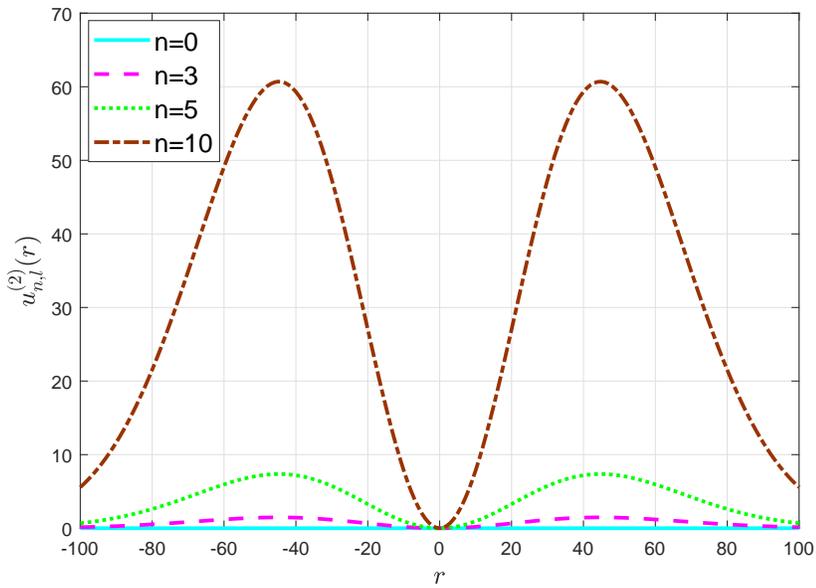

Figure 4: $u_{n,l}^{(2)}(r)$ versus $r$ with $l = 1$ and $\omega = 10^{-3}$.

# 4    The angular part solutions of Dirac equation

As mentioned before, for Hartmann and Ring-Shaped Oscillator potentials angular part of Dirac equation is [26-27]:

$$\frac{1}{\sin\theta}\frac{d}{d\theta}\left(\sin\theta\frac{d\Theta(\theta)}{d\theta}\right) - \left[\frac{m^2}{\sin^2\theta} + (\varepsilon + M^2c^4)f(\theta) - s\right]\Theta(\theta) = 0. \tag{4.1}$$

Assuming $\Theta(\theta) = \frac{H(\theta)}{\sin^{\frac{1}{2}}\theta}$, first-order differential term can be vanished from differential equation (4.1). This transformation can provide the condition that Schrödinger-like equation is accessible from Eq.(4.1). Considering the mentioned transformation, Eq.(4.1) can be converted to:

$$\frac{d^2H(\theta)}{d\theta^2} + \left[-\frac{(m^2 - \frac{1}{4})}{\sin^2\theta} - (\varepsilon + Mc^2)f(\theta) + \rho + \frac{1}{4}\right]H(\theta) = 0. \tag{4.2}$$

In comparison Eq.(4.2) with the following Schrödinger solvable equation ($\hbar = 2m = 1$) [28-29]:

$$\frac{d^2H(x)}{dx^2} + [E - V(x)]H(x) = 0, \tag{4.3}$$

Eq.(4.2) will be solvable according to different types of $f(\theta)$. It means that the solution of Eq.(4.3) for non-relativistic energy spectrum and different potentials will be expanded to Schrödinger-like equation obtained from Dirac equation. In this comparison relativistic parameters can be connected to non-relativistic parameters. Furthermore, it should be mentioned this method is useable for special functions of $f(\theta)$. Therefore, $f(\theta)$ functions that can be solved in this techniques as follows [26-27]:

$$f_1(\theta) = \frac{\gamma + \beta\cos\theta + \alpha\cos^2\theta}{\sin^2\theta}, \tag{4.4}$$

$$f_2(\theta) = \frac{\gamma + \beta\cos^2\theta + \alpha\cos^4\theta}{\sin^2\theta\cos^2\theta}, \tag{4.5}$$

$$f_3(\theta) = \gamma + \beta\cot\theta + \alpha\cot^2\theta, \tag{4.6}$$

where $\alpha$, $\beta$ and $\gamma$ are arbitrary constant values. In other words, above functions are solvable functions that are considered as Hartmann and Ring-Shaped Oscillator potentials. If $f_1(\theta)$ is substituted in Eq.(4.2) as:

$$\frac{d^2H(\theta)}{d\theta^2} + \left\{ \left[(-m^2 + \frac{1}{4}) - \eta(\gamma + \alpha)\right] \csc^2\theta - \eta\beta\csc\theta\cot\theta + \eta\alpha + \rho + \frac{1}{4} \right\} H(\theta) = 0, \quad (4.7)$$

where $\eta = \varepsilon + Mc^2$, Eq.(4.7) can be compared to the following Schrödinger equation ($\hbar = 2m = 1$) [28-29]:

$$\frac{d^2H(x)}{dx^2} + \left[-(\lambda^2 + s^2 - s)\csc^2 x + \lambda(2s - 1)\csc x\cot x + (s + n)^2\right]H(x) = 0. \quad (4.8)$$

The relations of parameters between non-relativistic solvable model and relativistic model will be obtained by comparing between Eqs.(4.7) and (4.8) as:

$$\eta\alpha + \rho + \frac{1}{4} = (s + n)^2, \quad (4.9)$$

$$\eta(\gamma + \alpha) + m^2 - \frac{1}{4} = \lambda^2 + s^2 - s, \quad (4.10)$$

$$\eta\beta = -\lambda(2s - 1). \quad (4.11)$$

The wave function that is related to Eq.(4.7) is written based on Jacobi polynomials $P_n^{(\mu,\nu)}(g(x))$ where $\mu > -1$, $\nu > -1$ and $n = 0, 1, 2, \dots$ . According to parameter definitions of $\mu = -\lambda + s - \frac{1}{2}$ and $\nu = \lambda + s - \frac{1}{2}$ in Jacobi polynomials, restrictions of $s$ and $\lambda$ parameters will be as $s > -\frac{1}{2}$ and $-(s + \frac{1}{2}) < \lambda < (s + \frac{1}{2})$.

The relation between non-relativistic energy and relativistic energy according to (4.9) causes $\rho$ separation constant is calculated as $\rho = (s + n)^2 - \alpha(\varepsilon + Mc^2) - \frac{1}{4}$, so that condition of $\rho + \frac{1}{4} \geq 0$ causes:

$$\varepsilon \leq \frac{1}{\alpha}(s + n)^2 - Mc^2. \quad (4.12)$$

Positive values may be provide for relativistic energy, if $\alpha > 0$. Non-normalized waved function that is associated to the solvable differential equation (4.8) is [28-29]:

$$H(x) = (1 - g)^{\frac{s-\lambda}{2}}(1 + g)^{\frac{s+\lambda}{2}}P_n^{(-\lambda+s-\frac{1}{2}, \lambda+s-\frac{1}{2})}(g(x)), \quad (4.13)$$

where $g(x) = \cos x$. Considering function (4.13), the wave function is obtained for differential equation (4.7) as follows:

$$H(\theta) = (1 - \cos\theta)^{\frac{s-\lambda}{2}}(1 + \cos\theta)^{\frac{s+\lambda}{2}}P_n^{(-\lambda+s-\frac{1}{2},\lambda+s-\frac{1}{2})}(\cos\theta). \qquad (4.14)$$

According to $\Theta(\theta) = \frac{H(\theta)}{\sin^{\frac{1}{2}}\theta}$, angular part of Dirac equation is gotten as:

$$\Theta^{(1)}(\theta) = 2^{s-1}(\sin\theta)^{s-\lambda-\frac{1}{2}}(\cos\theta)^{s+\lambda-\frac{1}{4}}P_n^{(-\lambda+s-\frac{1}{2},\lambda+s-\frac{1}{2})}(\cos\theta), \qquad (4.15)$$

where $-\frac{\pi}{2} \leq \theta \leq +\frac{\pi}{2}$. Wave function (4.15) will be zero, if $\theta$ variable is limited to the endpoints of interval, it means that when $\theta \to \mp\frac{\pi}{2}$ the wave function is restricted as $\Theta(\theta) \to 0$. Although establishing of mentioned physical situations and also to avoid divergence of wave function (4.15) at $\theta = 0$ will cause restriction of $\lambda$ and $s$ parameters changes to $-(s+\frac{1}{4}) < \lambda < (s-\frac{1}{2})$ for $s > \frac{3}{8}$.

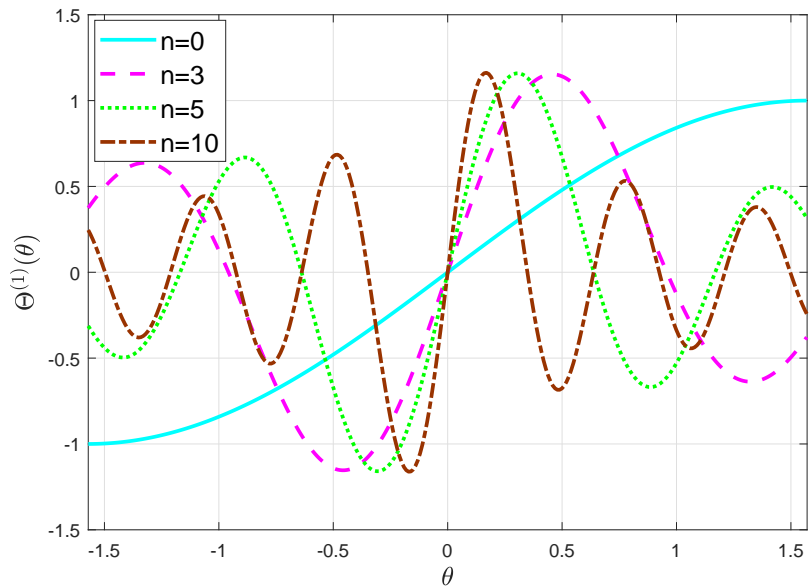

Figure 5: $\Theta^{(2)}(\theta)$ versus $\theta$ with $s = 1$ and $\lambda = -0.5$.

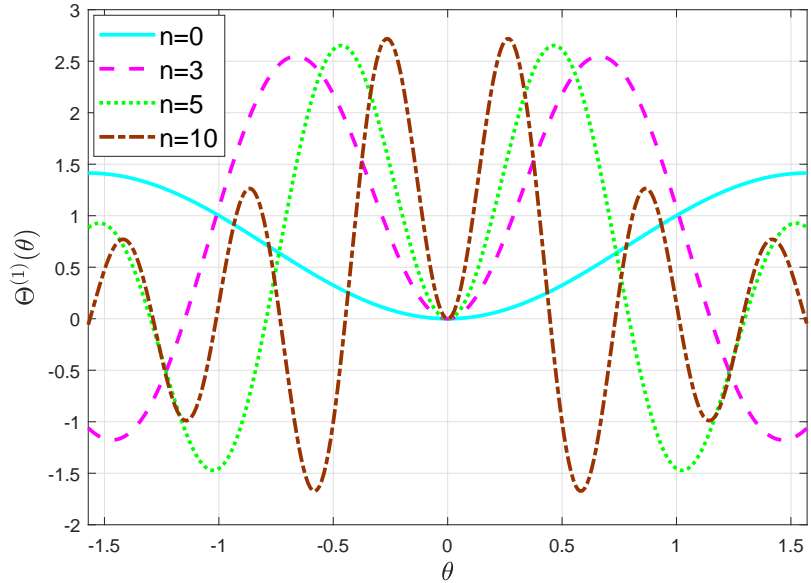

Figure 6: $\Theta^{(1)}(\theta)$ versus $\theta$ with $s = 1.5$ and $\lambda = -1$.

The illustrated technique can be expanded to other functions of $\theta$. For $f_2(\theta)$ function, angular part of Dirac equation has the following form:

$$\frac{d^2 H(\theta)}{d\theta^2} + \left\{ [(-m^2 + \frac{1}{4}) - \eta(\gamma + \beta + \alpha)] \csc^2 \theta - \eta\gamma \sec^2 \theta + \eta\alpha + \rho + \frac{1}{4} \right\} H(\theta) = 0. \quad (4.16)$$

Perfect differential solvable equation ($\hbar = 2m = 1$) that can be used for this method is [28-29]:

$$\frac{d^2 H(x)}{dx^2} + \left[ -\lambda(\lambda - 1) \csc^2 x - s(s - 1) \sec^2 x + (\lambda + s + 2n)^2 \right] H(x) = 0. \quad (4.17)$$

The following parameter relations are made by comparison between Eq.(4.16) and Eq.(4.17):

$$\eta\alpha + \rho + \frac{1}{4} = (\lambda + s + 2n)^2, \quad (4.18)$$

$$\eta(\gamma + \beta + \alpha) + m^2 - \frac{1}{4} = \lambda(\lambda - 1), \quad (4.19)$$

$$\eta\gamma = s(s - 1). \quad (4.20)$$

Since Jacobi polynomials $P_n^{(\mu,\nu)}$ is associated to Eq.(4.17) for $\mu > -1$ and $\nu > -1$, so $\lambda$ and $s$ will be restricted by $\lambda > -\frac{1}{2}$ and $s > -\frac{1}{2}$. Relation (4.18) confirms $\rho$ parameter as $\rho = (\lambda + s + 2n)^2 - \eta\alpha - \frac{1}{4}$ that can connect relativistic energy to perfect non-relativistic parameters. The condition of $\rho + \frac{1}{4} \geq 0$ creates the following range of relativistic energy spectrum:

$$\varepsilon \leq \frac{1}{\alpha}(\lambda + s + 2n)^2 - Mc^2. \tag{4.21}$$

If $\alpha$ parameter is considered as $\alpha > 0$, positive values may be gotten for relativistic energy spectrum.

Non-normalized wave function that is satisfied in differential equation (4.17), for $g(x) = \cos(2x)$ is [28-29]:

$$H(x) = (1 - g)^{\frac{\lambda}{2}}(1 + g)^{\frac{s}{2}} P_n^{(\lambda - \frac{1}{2}, s - \frac{1}{2})}(g(x)). \tag{4.22}$$

Function (4.22) can be expanded to differential equation (4.16) and considered as the exact solution of differential equation. The mentioned solution based on Jacobi polynomials is considered as:

$$H(\theta) = (1 - \cos 2\theta)^{\frac{\lambda}{2}}(1 + \cos 2\theta)^{\frac{s}{2}} P_n^{(\lambda - \frac{1}{2}, s - \frac{1}{2})}(\cos 2\theta), \tag{4.23}$$

so that angular part solution of Dirac equation can be constituted as:

$$\Theta^{(2)}(\theta) = 2^{\frac{\lambda + s}{2}}(\sin \theta)^{\lambda - \frac{1}{2}}(\cos \theta)^s P_n^{(\lambda - \frac{1}{2}, s - \frac{1}{2})}(\cos 2\theta), \tag{4.24}$$

where $-\frac{\pi}{4} \leq \theta \leq +\frac{\pi}{4}$. Wave function (4.24) will be always constant value at the endpoints of defined interval for $\theta$ variable, but divergence of the wave function at $\theta = 0$ converts the restriction of $\lambda$ parameter to $\lambda > \frac{1}{2}$. Therefore wave function (4.24) will be physically solution, if $s$ and $\lambda$ parameters are considered as $s > -\frac{1}{2}$ and $\lambda > \frac{1}{2}$.

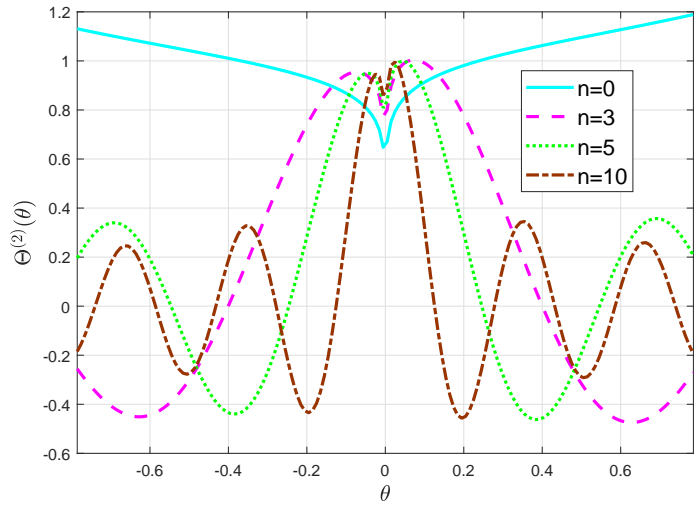

Figure 7: $\Theta^{(2)}(\theta)$ versus $\theta$ with $s = -0.2$ and $\lambda = 0.6$.

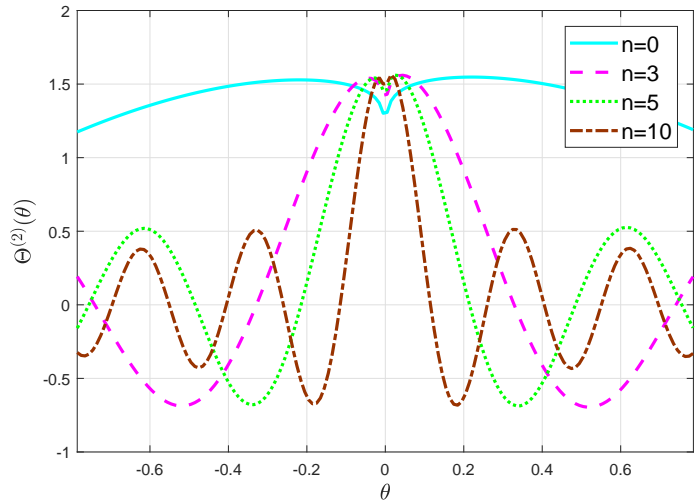

Figure 8: $\Theta^{(2)}(\theta)$ versus $\theta$ with $s = 1$ and $\lambda = 0.55$.

Another function of $f_3(\theta)$ can be also analyzed by this method because there is a non-relativistic solvable model which can be corresponded to this function as angular part solution of Dirac equation. Angular part of Dirac equation with $f_3(\theta)$ according to Eq.(4.2) is:

$$\frac{d^2 H(\theta)}{d\theta^2} + \left\{ [(-m^2 + \frac{1}{4}) - \eta\alpha]\csc^2\theta - \eta\beta\cot\theta - \eta(\gamma - \alpha) + \rho + \frac{1}{4} \right\} H(\theta) = 0. \quad (4.25)$$

For corresponding to non-relativistic solvable model, the following Schrödinger equation ($\hbar = 2m = 1$) is considered [28-29]:

$$\frac{d^2 H(x)}{dx^2} + \left[ -s(s+1)csc^2 x + 2\lambda\cot\theta + (s-n)^2 - \frac{\lambda^2}{(s-n)^2} \right] H(x) = 0. \quad (4.26)$$

The relativistic parameters in Eq.(4.25) connected to the non-relativistic parameters in Eq.(4.26) as follows:

$$\eta(\gamma - \alpha) - (\rho + \frac{1}{4}) = \frac{\lambda^2}{(s-n)^2} - (s-n)^2, \quad (4.27)$$

$$\eta\alpha + m^2 - \frac{1}{4} = s(s+1), \quad (4.28)$$

$$\eta\beta = -2\lambda. \quad (4.29)$$

In the assumed solvable model, the limit of $s$ and $\lambda$ parameters are considered as $s > n-1$ and $-i(s-n)(s-n+1) < \lambda < i(s-n)(s-n+1)$. By using relation (4.27) $\rho$ separation constant can be obtained as $\rho = \eta(\gamma - \alpha) + (s-n)^2 - \frac{\lambda^2}{(s-n)^2} - \frac{1}{4}$. The range of relativistic energy will be following form, if the condition $\rho + \frac{1}{4} \geq 0$ is considered:

$$\varepsilon \leq (\frac{1}{\alpha - \gamma})[(s-n)^2 - \frac{\lambda^2}{(s-n)^2}] - Mc^2. \quad (4.30)$$

If $\alpha > \gamma$ is considered, it will be possible to calculate positive value for relativistic energy spectrum. The following non-normalized wave function that is associated to Jacobi polynomials in the solvable model (4.26), for $g(x) = -i\cot x$, is [28-29]:

$$H(x) = (g^2 - 1)^{\frac{s-n}{2}} \exp\left( \frac{\lambda}{s-n}x \right) P_n^{(s-n+i\frac{\lambda}{s-n}, s-n-i\frac{\lambda}{s-n})}(g(x)). \quad (4.31)$$

The above non-normalized function based on Jacobi polynomials can be applied for differential equation (4.25) as follows:

$$H(\theta) = (-1)^{\frac{s-n}{2}}(\csc\theta)^{s-n} \exp\left( \frac{\lambda}{s-n}\theta \right) P_n^{(s-n+i\frac{\lambda}{s-n}, s-n-i\frac{\lambda}{s-n})}(-i\cot\theta). \quad (4.32)$$

Finally, angular part solution of Dirac equation that was called $\Theta(\theta)$ is gotten as:

$$\Theta^{(3)}(\theta) = (-1)^{\frac{s-n}{2}}(\csc\theta)^{s-n+\frac{1}{2}}\exp\left(\frac{\lambda}{s-n}\theta\right)P_n^{(s-n+i\frac{\lambda}{s-n},\,s-n-i\frac{\lambda}{s-n})}(-i\cot\theta), \qquad (4.33)$$

where $0 \le \theta \le \pi$. Restriction of $s$ parameter will be $s < n - \frac{1}{2}$, if the boundary situations are considered for wave function (4.33) in the endpoints of interval as $\theta \to 0$ and $\theta \to \pi$ also no divergence at $\theta = \frac{\pi}{2}$. Therefore, wave function (4.33) is physically acceptable by providing the range of $s$ and $\lambda$ parameters as $n - 1 < s < n - \frac{1}{2}$ and $-i(s-n)(s-n+1) < \lambda < i(s-n)(s-n+1)$.

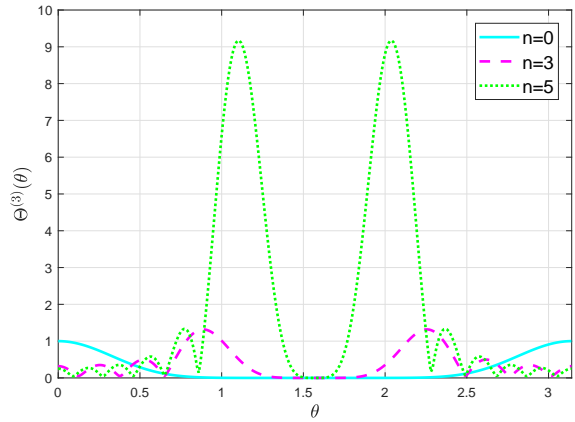

Figure 9: $\Theta^{(3)}(\theta)$ versus $\theta$ with $s = 9.5$ and $\lambda = 0.1i$.

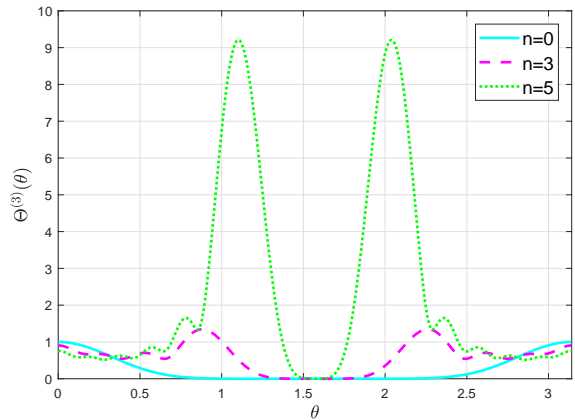

Figure 10: $\Theta^{(3)}(\theta)$ versus $\theta$ with $s = 9.5$ and $\lambda = 1i$.

# 5  Conclusion

The energy spectrum of bound states and spinor wave function of Dirac equation for Hartmann and Ring-Shaped Oscillator potentials have been calculated by comparing the mentioned relativistic models with non-relativistic systems. In radial and angular parts of Dirac equation, relativistic parameters and their restrictions have been investigated by considering the solutions of non-relativistic models that related to the problem and restrictions of non-relativistic parameters. By this method, spinor wave functions are associated to orthogonal polynomials such as Generalized Laguerre polynomials and Jacobi polynomials in radial and angular parts of Dirac equation, respectively.